\documentclass[12pt]{article}

\newcommand{\be}{\begin{equation}}\newcommand{\ee}{\end{equation}}
\newcommand{\barr}{\begin{array}}\newcommand{\earr}{\end{array}}
\newcommand{\bea}{\begin{eqnarray}}\newcommand{\eea}{\end{eqnarray}}
\newcommand{\nn}{\nonumber}\newcommand{\p}[1]{(\ref{#1})}
\newcommand{\lb}[1]{\label{#1}}
\newcommand\s{\scriptscriptstyle}

\newcommand\q{\quad}

\newcommand{\vp}{\varphi}
\newcommand{\bvp}{{\bar\varphi}}

\newcommand{\bnu}{{\bar\nu}}
\newcommand{\bmu}{{\bar\mu}}

\newcommand\cE{{\cal E}}

\newcommand\cL{{\cal L}}

\newcommand{\da}{{\dot{\alpha}}}
\newcommand{\db}{{\dot{\beta}}}

\newcommand\ab{{\alpha\beta}}

\newcommand\adb{{\alpha\db}}
\newcommand\ada{{\alpha\da}}
\newcommand\bdb{{\beta\db}}

\newcommand\padb{\partial_\adb}

\newcommand\B{{\s B}}

\newcommand\D{{\s D}}
\newcommand\I{{\s I}}

\begin{document}
\begin{center}
{\bf NEW APPROACH TO NONLINEAR ELECTRODYNAMICS: \break
DUALITIES AS SYMMETRIES OF INTERACTION}\\
\vspace{0.5cm}
{\bf E.A. Ivanov, B.M. Zupnik}\\
\end{center}
\vspace{0.2cm}

\centerline{\it Joint Institute for Nuclear Research,}
\centerline{\it Dubna, Moscow Region, 141980 Russia}
\vspace{0.2cm}

\begin{abstract}
\noindent We elaborate on the duality-symmetric nonlinear electrodynamics
in a new formulation with auxiliary tensor fields. The Maxwell field 
strength appears only in bilinear terms of the corresponding generic 
Lagrangian, while the self-interaction is presented by a function $E$ 
depending on the auxiliary fields. Two types of dualities  inherent in the
nonlinear electrodynamics models admit a simple off-shell characterization 
in terms of this function. In the standard formulation, the continuous 
$U(1)$ duality symmetry is nonlinearly realized on the Maxwell field 
strength. In the new setting, the same symmetry acts as linear $U(1)$ 
transformations of the auxiliary field variables. The nonlinear $U(1)$  
duality condition proves to be equivalent to the linear $U(1)$ invariance
of the self-interaction $E$. The discrete self-duality (or self-duality 
by Legendre transformation) amounts to a weaker reflection symmetry of $E$.
For a class of duality-symmetric Lagrangians we introduce an alternative
representation with the auxiliary scalar field and find new explicit
examples of such systems.
\end{abstract}
\vspace{0.5cm}

\setcounter{equation}0
\section{Introduction}
It is well known that the on-shell $SO(2)$ ($U(1)$) duality symmetry (or
self-duality) of Maxwell equations can be generalized to the whole class of
the nonlinear electrodynamics models, including the famous Born-Infeld
theory. The condition of  $SO(2)$ duality can be formulated as a nonlinear
differential constraint on the  Lagrangians of these models \cite{GZ,GR,KT}.
Using a non-analytic change of  basic field variables of the Lagrangian
the $SO(2)$ duality condition can be transformed to the well-known
Courant-Hilbert equation \cite{PS,GZ2,HKS}. Recently, it has been observed 
that the requirement of analyticity of the initial Lagrangian implies an 
additional algebraic constraint which selects the proper subclass of 
solutions of this Courant-Hilbert equation \cite{BC}.

In this paper we elaborate on another approach to the $U(1)$
duality-symmetric Lagrangians, in which the manifest analyticity is 
guaranteed at each step.\footnote{A preliminary version of this approach 
was presented in \cite{IZ,IZ1}.} It makes use of the auxiliary tensor 
fields. The starting point is the generic nonlinear electrodynamics 
Lagrangian
\be
L(F^2,\bar F^2) = -{1\over 2}(F^2 + \bar F^2) + L^{int}(F^2, \bar F^2)\,,
\lb{1}
\ee
where $F^2 = F_{\alpha\beta}F^{\alpha\beta}$,$\bar F^2 = 
\bar F_{\dot\alpha\dot\beta}\bar F^{\dot\alpha\dot\beta}$ and 
$F_{\alpha\beta},\bar F_{\dot\alpha\dot\beta}$ are the mutually conjugated 
$(1,0)$ and $(0,1)$ components of the Maxwell field strength in the 
two-component spinor notation. Its new representation involves, apart from 
the Maxwell field strength, also unconstrained auxiliary symmetric bispinor 
(tensor) fields $V_{\alpha\beta},\bar V_{\dot\alpha\dot\beta}$ and their 
squares $V^2 \equiv \nu$ and $\bar V^2 \equiv \bar\nu$
\be
\cL(F,V) = {1\over 2}(F^2 + \bar F^2) +(\nu +\bnu)- 2 (V\cdot F + 
\bar V\cdot F) + E(\nu, \bnu)\,,\lb{2}
\ee
where $E(\nu, \bnu)$ codifies the entire self-interaction. The generic 
Lagrangian \p{1} is recovered as a result of eliminating the auxiliary 
fields in \p{2} by their algebraic equations of motion. The basic advantage 
of this novel representation for $U(1)$ duality-symmetric systems is 
related to its following remarkable feature. In contradistinction to 
nonlinear $U(1)$ duality transformations of $F^2,\bar F^2$, the 
transformations of the new auxiliary variables are {\it linear}. As a 
consequence, the $SO(2)$ duality condition is linearized and can be 
explicitly solved in this new setting. The general Lagrangian solving
this constraint is specified by the interaction term  $E^{ds}(\nu,\bar\nu) 
=\cE( \nu\bar\nu)$ which includes only the $U(1)$-invariant scalar
combination of the auxiliary fields $\nu\bar\nu = V^2\bar V^2$ as an 
argument. More general nonlinear electrodynamics Lagrangians respecting 
the so-called discrete self-duality (or duality by Legendre transformation) 
also admit a simple off-shell characterization in terms of the function 
$E$. In this case it should be even, $E(\nu,\bar\nu) =E(-\nu, -\bar\nu)$, 
and otherwise arbitrary.

The paper is organized as follows. In Sect. \ref{A} we give a brief 
account of the continuous and ``discrete'' dualities in nonlinear 
electrodynamics in the conventional approach. A novel representation of 
the appropriate Lagrangians via bispinor auxiliary fields is discussed in 
Sect. \ref{B}. An explicit solution of the algebraic equations of motion 
relating the initial and auxiliary variables can be immediately found only 
for a restricted class of the interaction functions $\cE(\nu\bnu)$, e.g. 
for the text-book case of Born-Infeld theory (though their perturbative 
solution always exists). In order to construct new explicit examples of 
duality-symmetric models we introduce, in Sect. \ref{D}, an alternative 
representation for the important subclass of interactions $L^{int}(F^2, 
\bar F^2)$, namely, those containing terms of the 4th order in the field 
strengths. This representation makes use of a different linearly
transforming auxiliary scalar variable $\mu$ which is related to the 
variable $\nu = V^2$ by a sort of Legendre transformation. The ansatz for 
the appropriate class of solutions of the $U(1)$ duality condition
contains an invariant analytic function $I(\mu\bmu)$, and the  expression
for the corresponding Lagrangians in the $\mu$-representation is 
parametrized by this function. The algebraic equation for the auxiliary 
variable $\mu$ can be explicitly solved in terms of the initial variables 
$ F^2, \bar F^2$ for a wide class of the functions $I(\mu\bar\mu)$.
Explicit examples of duality-symmetric analytic Lagrangians, including 
the Born-Infeld Lagrangian and some new ones constructed here for the 
first time, are collected in Sect. \ref{C}.

\setcounter{equation}0
\section{Dualities in nonlinear electrodynamics\lb{A}}

We start by recapitulating the basic facts about nonlinear 4D 
electrodynamics models which reveal duality properties and include the 
free Maxwell theory and Born-Infeld theory as particular cases. Detailed 
motivations why such models are of interest to study can be found e.g. in 
\cite{KT}.

\subsection{Continuous on-shell $SO(2)$ duality}
In the two-component spinor  notation, the Maxwell field strengths are
defined by
\bea
&&F_\alpha^\beta(A)\equiv {1\over4}(\sigma^m)_\adb
(\bar\sigma^n)^{\db\beta}F_{mn}={1\over4}(\partial_\alpha^\db A_\bdb
+\partial_\beta^\db A_\adb)~,\lb{defF}\\
&&\bar{F}_\da^\db(A)\equiv\overline{F_\alpha^\beta(A)}~,\q F_{mn}=
\partial_m A_n-\partial_nA_m~,\nn
\eea
where $\sigma^m, \bar\sigma^n$ are the Weyl matrices of the group 
$SL(2,C)$, $\padb=(\sigma^m)_\adb \partial_m$ and  $A_\adb=(\sigma^m)_\adb 
A_m$ is the corresponding vector gauge potential. Below we shall sometimes 
treat $F_\ab$  $(\bar{F}_{\da\db})$ as independent variables, without 
assuming them to be expressed through $A_m$.

Let us introduce the Lorentz-invariant complex variables
\be
 \varphi\equiv F^2 = F^\ab F_\ab~,\quad \bar\varphi\ = \bar F^2  =
\bar{F}^{\da\db}\bar{F}_{\da\db}~.
\lb{compl}
\ee
Two independent real invariants which one can construct out of the Maxwell 
field strength in the standard vector notation take the following form in 
these complex variables:
\bea
F^{mn}F_{mn}=2( \vp+\bvp)~,\quad {1\over2}\varepsilon^{mnpq}F_{mn}F_{pq} 
=-2i(\vp-\bvp)~. \lb{vecsp}
\eea

It will be convenient to deal with dimensionless $F_{\alpha\beta},
\bar{F}_{\dot\alpha\dot\beta}$ and $\vp, \bvp$, introducing a coupling
constant $f$, $[f] = 2$. Then the generic nonlinear Lagrangian
can be represented as
$$
f^{-2}L(\vp, \bvp)\,,
$$
where
\be
L(\vp,\bvp) =
-{1\over2}(\vp+\bvp)+L^{int}(\vp,\bvp)
\lb{nonl}
\ee
and {\it the real analytic} self-interaction $L^{int}(\vp, \bvp)$ collects
all possible higher-order terms   $\vp^k \bvp^m~, (k+m)\geq 2$. This
analyticity requirement rules out, for instance, terms with radicals
of the type $\sqrt\vp$ or $\sqrt{\vp\pm\bvp}$.

We shall use the following notation for the derivatives of the Lagrangian
$L(\vp,\bvp)$ \footnote{In these and some subsequent relations it is
assumed that the functional argument $F$ stands for both $F_\ab$ and
$\bar{F}_{\da\db}$; we hope that this short-hand notation will not give 
rise to any confusion.}
\bea
&&P_\ab(F)\equiv i\partial L/\partial F^\ab=2iF_\ab
 L_\vp~,\lb{defP}\\
&&L_\vp= \partial L/\partial \vp~,\q L_\bvp= \partial L/ \partial
\bvp\nn
\eea
and for the bilinear combinations of them
\bea
\pi\equiv P^2=P^\ab P_\ab=-4\vp(L_\vp)^2~,\quad
\bar\pi = \bar P^2=
\bar{P}^{\da\db}\bar{P}_{\da\db}=-4\bvp(L_\bvp)^2~.\lb{Pvar}
\eea
In the vector notation, the same quantities read
\bea
\widetilde{P}^{mn}\equiv{1\over2}\,\varepsilon^{mnpq}P_{pq}=
2 \partial L/\partial F_{mn}\,, \quad
{i\over2}P_{mn}\widetilde{P}^{mn}=\pi-\bar{\pi}~.\nn
\eea

The nonlinear equations of motion have the following form in the
spinor notation:
\bea
&& \partial_\alpha^\db \bar P_{\da\db}(F)
-\partial^\beta_\da P_\ab(F)= 0~.\lb{BIeq}
\eea

These equations, together with the Bianchi identities
\bea
&&
\partial_\alpha^\db \bar F_{\da\db}-\partial^\beta_\da F_\ab
= 0~, \lb{Bian}
\eea
constitute a set of first-order equations in which one can treat $F_\ab$
and $\bar{F}_{\da\db}$ as {\it unconstrained} conjugated variables.

This set is said to be duality-symmetric if the Lagrangian $L(\vp, \bvp)$
satisfies certain nonlinear condition \cite{GZ,GR,KT}. The precise form
of this $SO(2)$ duality condition is as follows
\bea
&&F^2+P^2-\bar F^2-\bar P^2\equiv {i\over4}\varepsilon^{mnpq}
(F_{mn} F_{pq}+P_{mn} P_{pq})
\nn\\
&&= \vp + \pi -\bvp - \bar\pi =\vp - \bvp - 4\,[\vp(L_\vp)^2 - 
\bvp(L_\bvp)^2]=0\,.\lb{sdI}
\eea
To clarify the meaning of \p{sdI}, let us define the nonlinear 
transformations
\bea
&& \delta_\omega F_\ab=\omega\,P_\ab(F) = 2i\,\omega\,F_\ab L_\vp~,
\lb{Ftrans}\\
&&\delta_\omega\vp=4i\omega\vp L_\vp\lb{vptrans}
\eea
where $\omega$ is a real parameter. Then eq. \p{sdI} ensures that this 
transformation constitutes a nonlinear realization of the $SO(2)$ group. 
Indeed, given \p{sdI}, $F_{\alpha\beta}$ and $P_\ab(F)$ form
an $SO(2)$ vector
\be
\delta_\omega P_\ab(F)
=-\omega F_\ab~ \lb{Ptrans1}~.
\ee
The set of equations \p{BIeq}, \p{Bian} and the  constraint \p{sdI} 
itself are clearly invariant under these transformations. Thus they are 
an obvious generalization of the $SO(2)$ duality transformation in the 
Maxwell theory:
\be
 \delta_\omega  F_\ab=-i\omega\, F_\ab~, \q \delta_\omega
\bar{F}_{\da\db}= i\omega\,\bar{F}_{\da\db}~,
\ee
which is a symmetry of the vacuum Maxwell equation
$\partial^\beta_\da F_\ab=0\,$.

Using \p{sdI}, the following important relations can be derived:
\be
\delta_\omega L =i\omega(\vp-\bvp)
~,\q\delta L_\vp={i\over2}\omega-2i\omega L_\vp^2~.
\lb{delta1}
\ee
It should be pointed out that these transformations make sense only 
on the mass shell defined by eqs. \p{BIeq}, \p{Bian}.

The general solution of the $SO(2)$ duality condition \p{sdI} has been
considered earlier in Refs.\cite{GR,PS,GZ2,HKS,BC}. Using the nonanalytic
change of variables
\be
p={1\over4}(\vp+\bvp)+{1\over2}\sqrt{\vp\bvp}~,\q q={1\over4}(\vp+\bvp)-
{1\over2}\sqrt{\vp\bvp}
\ee
one can cast equation \p{sdI} in the form of the well-known
Courant-Hilbert equation
\be
\cL_p\cL_q=1~.\lb{CHeq}
\ee
The general solution of this equation is parametrized by a real analytic
function $v(s)\,$,
\bea
&&{\bf L}(p,q)=v(s)+\frac{2p}{v^\prime(s)}~,\q q=s+
\frac{p}{[v^\prime(s)]^2}\lb{nasd}
\eea
and it is completely specified by an algebraic equation for the auxiliary
variable $s\,$. The authors of Ref. \cite{BC} have shown that the natural 
requirement of analyticity of the Lagrangian with respect to the initial 
variables $\vp,\bvp$ can be rephrased as the additional constraint on the 
function $\Psi(s)=-s[v^\prime(s)]^2$
\be
\Psi[\Psi(s)]=s~.\lb{addit}
\ee
The perturbative analysis shows that the whole class of duality-symmetric
analytic solutions ${\bf L}[p(\vp,\bvp),q(\vp,\bvp)]$  exists. However,
the only solution explicitly worked out so far is the familiar
Born-Infeld example. Nonphysical solutions of eq.\p{CHeq} contain 
nonanalytic terms  $\sqrt{\vp\bvp}\,$ (see, e.g. \cite{HKS}).

In Sect. 3 and 4 we shall discuss two complementary approaches to solving 
the $SO(2)$ duality equation which guarantee analyticity and covariance 
of solutions at each stage of calculations. Based on this, in Sect. 5 we 
shall present several new examples of duality-symmetric Lagrangians which 
meet the analyticity criterion.

It is worth pointing out once more that the $SO(2)$ duality transformations 
in the standard setting described above cannot be realized on the vector
potential $A_m$; they provide a symmetry between the equations of motion
and Bianchi identity and as such define {\it on-shell} symmetry. The
manifestly $SO(2)$ duality-invariant off-shell Lagrangians can be 
constructed in the formalism with additional vector and auxiliary fields 
\cite{PST}. We are planning to discuss a relation to this extended 
formalism elsewhere.

The Lagrangian $L(\vp,\bvp)$ satisfying \p{sdI} is not invariant with 
respect to transformation \p{Ftrans}. Yet one can construct, out of $\vp$ 
and $\bvp\,$, the $SO(2)$ invariant function
\bea
I(\vp,\bvp)\equiv L+{i\over2}(F\cdot P-\bar{F}\cdot\bar{P}) =
L-\vp L_\vp-\bvp L_\bvp~,\lb{invar}
\eea
where $F\cdot P= F^\ab P_\ab $. However, $I(\vp,\bvp)$
starts with the 4-th order term $\vp\bvp$, so this invariant cannot be
interpreted as a Lagrangian.

Finally, notice that, given some $L^{ds}(\vp, \bvp)$ obeying \p{sdI},
the following Lagrangian related to $L^{ds}$ by the simple rescaling
\be
L^{ds}(\vp,\bvp)~\Rightarrow~rL^{ds}(r^{-1}\vp,r^{-1}\bvp)~,\lb{rescal}
\ee
with $r\neq 0$ being an arbitrary real number, also obeys \p{sdI} and
so yields a duality-symmetric model. Clearly, rescaling the coupling 
constant as $f^2\rightarrow |r|f^2$ and properly rescaling 
$F_{\alpha\beta}, \bar F_{\dot\alpha\dot\beta}$, one can always choose 
$|r| =1$, so only the sign of $r$ actually matters in \p{rescal}.
Thus
\be
L^{(-)}(\vp,\bvp)=-L^{ds}(-\vp,-\bvp)~\lb{mirr}
\ee
gives a non-equivalent duality-symmetric Lagrangian for each given 
$L^{ds}$. In Sect.\ref{C} we shall consider this transformation for the 
Lagrangian of the Born-Infeld theory.

\subsection{Self-duality by Legendre transformation}
To explain what the ``discrete duality'' means we shall need a first-order 
representation of the action corresponding to the Lagrangian \p{nonl}. It 
is such that the Bianchi identities \p{Bian} are implemented in the action 
with a Lagrange multiplier and so $F_\ab, \bar F_{\da\db}$ are 
unconstrained complex variables off shell. This form of the action is 
given by
\be
{1\over f^2}\int d^4x L^\D(F,F^\D)={1\over f^2}
\int d^4x[L(\vp,\bvp)+i(F\cdot F^\D-\bar{F}\cdot\bar{F}^\D)]~,\lb{1stor}
\ee
where
\be
 F^\D_{\alpha\beta}\equiv{1\over4}(\partial_{\alpha}^\db A^\D_{\beta\db}
+\partial_{\beta}^\db A^\D_{\alpha\db})~.\lb{PD}
\ee
Varying with respect to the Lagrange multiplier $A^\D_\adb\,$, one obtains 
just the Bianchi identities for $F_\ab, \bar F_{\da\db}$ \p{Bian}. Solving 
them in terms of the gauge potential $A_{\alpha\dot\beta}$ and substituting 
the result into \p{1stor}, we come back to \p{nonl}. On the other hand, 
the multiplier $A^\D_\adb$ is defined up to the standard Abelian gauge
transformation, which suggests interpreting $A^\D_\adb$ and
$F^\D_{\alpha\beta}$ as the {\it dual} gauge potential and gauge field
strength, respectively. Using the algebraic equations of motion for the
variables $F_\ab, \bar F_{\da\db}\,$, one can express the action \p{1stor}
in terms of $F^\D_{\alpha\beta}, \bar F^\D_{\da\db}\,$. If the resulting
action has the same form as the original one in terms of $F_\ab(A),
\bar F_{\da\db}(A)$, the corresponding model is said to enjoy the 
discrete duality. This sort of duality should not be confused with the 
on-shell continuous $SO(2)$ duality discussed earlier. However, as we 
shall see soon, any $L(\vp, \bvp)$ solving the constraint \p{sdI} defines 
a system possessing the discrete duality. The inverse statement is not 
generally true, so the class of nonlinear electrodynamics actions
admitting $SO(2)$ duality of equations of motion forms a subclass in the
variety of actions which are duality-symmetric in the ``discrete'' sense.

Let us elaborate on this in some detail. The dual picture is achieved by
varying \p{1stor} with respect to the independent variables $F_\ab,
\bar{F}_{\da\db}$, which yields the equation
\be
F^\D_{\alpha\beta}=i \partial L/\partial F^\ab \equiv P_\ab(F) =
2iF_\ab L_\vp~,
\label{FDF}
\ee
where $P_\ab(F)$ is the same as in \p{defP}. Substituting the solution of
this algebraic equation, $F_\ab = F_\ab(F^\D)$, into \p{1stor} gives us
the dual Lagrangian $L^\prime(F^\D)$
\be
L^\prime(\varphi^\D, \bar\varphi^\D) \equiv
L^\D[F(F^\D), F^\D]~,
\lb{hatL}
\ee
where $\varphi^\D \equiv F^{\D\,\ab}F^\D_{\ab} = \pi(F)$ and $\pi, 
\bar\pi$ were defined in \p{Pvar}. Then the discrete self-duality 
defined above amounts to the condition
\be
L^\prime (\varphi^\D, \bar\varphi^\D) = L(\varphi^\D, \bar\varphi^\D)~, 
\lb{discr1}
\ee
or, equivalently, to
\be
L^\prime (\pi, \bar\pi) = L(\pi, \bar\pi)~. \lb{discr}
\ee

Using \p{FDF} and its conjugate, as well as the definitions \p{1stor},
\p{hatL}, one can explicitly check the property
\be
 F_\ab=-i \partial L^\prime(\varphi^\D, \bar\varphi^\D)/\partial 
 F^{\D\,\ab} \quad (\mbox{and c.c.})~.
\ee
Due to this relation, and keeping in mind the inverse one \p{FDF},
one can treat the equation
\bea
L^\prime(P^2, \bar P^2)
=L(F^2,\bar F^2)+i(F\cdot P-\bar{F}\cdot\bar{P})
=L(\vp,\bvp)-2\vp L_\vp-2\bvp L_\bvp \lb{stleg}
\eea
as setting the Legendre transforms $L\leftrightarrow L^\prime$ between 
two functions of complex variables. Thus the discrete duality \p{discr1}, 
\p{discr} can be equivalently called ``self-duality by Legendre 
transformation''.

On the level of equations of motion \p{BIeq} and \p{Bian}, the discrete
self-duality \p{discr} can be equivalently defined as their invariance with 
respect to the special finite $SO(2)$ transformation $\Lambda$
\bea
&& F_\ab\rightarrow \Lambda F_\ab=P_\ab~,\q P_\ab\rightarrow \Lambda P_\ab
=-F_\ab \,.\lb{discrSO}
\eea
This invariance is manifested in the following on-shell transformation 
properties of the Lagrangian and its derivative
\bea
&&\Lambda L(\vp,\bvp)=L(\vp,\bvp)+iP\cdot F-i\bar P\cdot\bar F\equiv
L(\pi,\bar\pi)\,,\lb{Lambda}\\
&&\Lambda L_\vp={1\over4}L_\vp^{-1}~.\nn
\eea

Let us show  that the $SO(2)$ duality condition \p{sdI} indeed guarantees
the discrete duality \p{discr}. The simplest proof of this statement (see 
e.g. \cite{KT}) makes use of the special $SO(2)$ transformation $\Lambda$, 
eq. \p{discrSO}, and the invariance of function \p{invar} under the global 
version of the general $SO(2)$ transformations \p{Ftrans}
\be
\Lambda I(\vp,\bvp)\equiv L(\pi,\bar\pi)-{i\over2}F\cdot P+
{i\over2}\bar{F}\cdot\bar{P}=I(\vp,\bvp)\,.\lb{discinvar}
\ee
Comparing this relation with \p{stleg}, we arrive at the condition
\p{discr}. Clearly, the $\Lambda$-invariance of $I(\vp,\bvp)$ is a weaker 
condition than its $SO(2)$ invariance, so the Lagrangians revealing the 
property of $SO(2)$ duality form a subclass of those which are self-dual 
in the discrete sense.

\setcounter{equation}0
\section{The nonlinear electrodynamics and
dualities \break revisited\lb{B}}
\subsection{A new setting for Lagrangians of nonlinear electrodynamics}
The recently constructed $N=3$ supersymmetric extension of the Born-Infeld
theory \cite{IZ} suggests a new representation for the actions of
nonlinear electrodynamics discussed in the previous Section.

The infinite-dimensional off-shell $N=3$ vector multiplet contains gauge 
field strengths \p{defF}  and auxiliary fields $V_\ab$ and
$\bar{V}_{\da\db}\,$.

The gauge field part of the off-shell super $N=3$ Maxwell component 
Lagrangian is \footnote{In the rest of the paper we put the overall 
coupling constant $f$ equal to 1.}
\be
\cL_2(V,F)=\nu+ \bnu- 2\,(V\cdot F+\bar{V}\cdot\bar{F})
+{1\over2}(\vp+\bvp)~, \lb{auxfree}
\ee
where
\bea
&&\nu\equiv V^2=V^\ab V_\ab~,\q\bnu\equiv\bar V^2=\bar{V}^{\da\db}
\bar{V}_{\da\db}~,\nn\\
&& V\cdot F\equiv V^\ab F_\ab~,\q\bar{V}\cdot\bar{F}\equiv 
\bar{V}^{\da\db}\bar{F}_{\da\db}\,.
\eea
Eliminating $V^\ab$ by its algebraic equation of motion,
\be
 V^\ab = F^\ab~, \q \bar V^{\da\db} = \bar F^{\da\db}~,\lb{free1}
\ee
we arrive at the free Maxwell Lagrangian
\be
 L_2(F) = - {1\over2}(\vp + \bvp)~. \lb{maxw1}
\ee

Our aim will be to find a nonlinear extension of the free Lagrangian 
\p{auxfree}, such that this extension becomes the generic nonlinear 
Lagrangian $L(F^2,\bar{F}^2)$, eq. \p{nonl}, upon eliminating the 
auxiliary fields $V_{\alpha\beta}, \bar{V}_{\dot\alpha\dot\beta}$ by
their algebraic ({\it nonlinear}) equations of motion.

By Lorentz covariance, the off-shell $(F,V)$-representation of the 
nonlinear Lagrangian \p{nonl} has the following general form:
\be
 \cL[V,F(A)] = \cL_2[V,F(A)] + E(\nu,\bnu)~, \lb{legact}
\ee
where $E$ is a real analytic function of two variables which encodes
self-interaction. Varying the action with respect to $V_\ab\,$, we derive 
the analytic relation between $V$ and $F(A)$ in this formalism
\be
 F_\ab(A) = V_\ab(1+ E_\nu) \quad (\mbox{and c.c.})~, \lb{FV}
\ee
where $E_\nu\equiv\partial E(\nu,\bnu)/\partial\nu$. The corresponding
algebraic relations between the scalar functions are
\be
\vp=\nu(1+E_\nu)^2~,\q F\cdot V=\nu(1+E_\nu)~.\lb{algsc}
\ee

The relation \p{FV}  can be used to eliminate the auxiliary variable 
$V_\ab$ in terms of $F_\ab$ and $\bar F_{\da\db}$, $V_\ab~\Rightarrow~
V_\ab[F(A)]$ (see eq. \p{VF} below). The natural restrictions on the 
interaction function $E(\nu, \bnu)$ are
\bea
&& E(0,0)=0~,\q E_\nu(0,0)= E_\bnu(0,0) = 0~.\lb{stcond}
\eea
They mean that the $(\nu, \bnu)$-expansion of $E(\nu, \bnu)$ does not 
contain constant and linear terms. Clearly, given some analytic 
interaction Lagrangian $L^{int}(\vp, \bvp)$ in \p{nonl}, one can pick up 
the appropriate function $E(\nu, \bnu)$, such that the elimination of 
$V_\ab,\bar V_{\da\db}$ by \p{FV} yields just this self-interaction.
Thus \p{legact} with an arbitrary (non-singular) interaction function $E$
is an alternative form of generic nonlinear electrodynamics Lagrangian 
\p{nonl}. The second equation of motion in this representation, obtained 
by varying \p{legact} with respect to $A_\ada\,$, has the form
\be
 \partial^\beta_\da[F_\ab(A)-2V_\ab]+\mbox{c.c.}=0~.\lb{FV3}
\ee
After substituting $V_\ab = V_\ab[F(A)]$ from \p{FV}, eq. \p{FV3} becomes
the dynamical equation for $F_\ab(A), \bar F_{\da\db}(A)$ corresponding
to the generic Lagrangian  \p{nonl}. Comparing \p{FV3} with \p{BIeq}
yields the important relation
\be
 P_\ab(F) = i\left[ F_\ab - 2V_\ab(F) \right]~, \lb{imprel}
\ee
where $P_\ab(F)$ was defined in \p{defP}.

Let us elaborate in more detail on how the $(F, V)$-representation of the 
nonlinear electrodynamics Lagrangians is related to the original 
``minimal'' one \p{nonl}. The general solution of the algebraic equation 
\p{FV} for $V_\ab$ can be written as
\be
 V_\ab(F)=F_\ab G(\vp,\bvp)\lb{VF}~.
\ee
The relation of the transition functions $G,\bar G$ to $E(\nu,\bnu)$
follows from  eq. \p{FV}
\be
 G^{-1}=1+E_\nu~,\q\bar{G}^{-1}=1+E_\bnu~. \lb{GE}
\ee
Eq.\p{VF} gives us the relations
\bea
&&\nu=\vp G^2~,\q\bnu=\bvp\bar{G}^2\lb{nuvp}~, \\
&&V(F)\cdot F=\vp G~,\q\bar{V}(F)\cdot\bar F=\bvp\bar{G}~,\lb{VF2}
\eea
which, taking into account \p{GE}, coincide with \p{algsc}.

The transition function $ G(\vp, \bvp)$ can be found from the basic
requirement that  \p{legact} coincides with the initial nonlinear
action after eliminating $V_\ab, \bar V_{\dot\alpha\dot\beta}$:
\bea
&& \cL[V(F),F]=L(\vp,\bvp)~.\lb{EL1}
\eea
Using eqs. \p{imprel}, \p{VF2} and the definition \p{defP}, it is easy 
to obtain the simple expression for the transition function in terms of 
the Lagrangian \p{nonl}
\bea
&& G(\vp, \bvp)= {1\over 2}-L_\vp~.
\lb{exprG}
\eea
A useful corollary of this formula and of eqs. \p{nuvp}, \p{GE} is
\be
 \nu E_\nu={1\over4}\vp(1-4L^2_\vp)~. \lb{corol}
\ee

Given a fixed $L(\vp, \bvp)$, one can express $\vp,\bvp$ (and then $G, 
\bar G$) in terms of $\nu, \bnu$ from eqs. \p{nuvp}, \p{exprG} and restore 
the explicit form of $E(\nu, \bnu)$ from \p{legact}, \p{auxfree},
\bea
E = L(\vp, \bvp) -{1\over2}(\vp+\bvp)-\nu-\bnu+2(\vp G + \bvp \bar G)~, 
\lb{ELrel}
\eea
via the substitution $\vp, \bvp \rightarrow \vp(\nu, \bnu), \bvp(\nu, 
\bnu)$. Conversely, given $E(\nu, \bnu)$, one can restore $L(\vp, \bvp)$,
by expressing $\nu$ through $\vp, \bvp$ from the first of eqs. \p{algsc}.
In practice, finding such explicit relations is a rather complicated task
(see Sect. \ref{C}).

\subsection{Duality symmetries as invariance of self-interaction}
So far we did not discuss dualities in the $(F,V)$-representation. A link
with the consideration in the previous Section is established by eq.
\p{imprel} which relates the functions $P_\ab(F)$ and $V_\ab(F)$.

Using this identification, the realization of the $SO(2)$ duality 
transformations \p{Ftrans}, \p{Ptrans1} on independent variables $F_\ab$ 
and $V_\ab$ is easily found to be
\bea
&&\delta_\omega V_\ab=-i \omega V_\ab~,\lb{Vtrans}\\
&&\delta_\omega F_\ab =i \omega [F_\ab -2V_\ab]~.\nn
\eea
We see that, before effecting the algebraic equation \p{FV} which 
expresses $V_{\alpha\beta}$ in terms of $F_{\alpha\beta}$ and 
$\bar F_{\dot\alpha\dot\beta}\,$, $SO(2)$ duality symmetry is realized
{\it linearly}.

Next, substituting \p{imprel} into the $SO(2)$ duality condition \p{sdI} 
and making use of eq. \p{corol}, we find
\be
{1\over4}\vp(1-4L^2_\vp)-{1\over4}\bvp(1-4L^2_\bvp)=\nu E_\nu-
\bnu E_\bnu=0~. \lb{newsd}
\ee
Thus passing to the $(F,V)$-representation allows one to rewrite the
nonlinear differential equation \p{sdI} as a {\it linear} differential
equation for the function $E(\nu, \bnu)$. It is important to emphasize that 
the new form \p{newsd} of the constraint \p{sdI} admits a transparent 
interpretation as the condition of invariance of $E(\nu,\bnu)$ under the
$U(1)$ transformations \p{Vtrans}
\bea
&& \delta_\omega E = 2i\omega (\bnu E_\bnu - \nu E_\nu) = 0~.
\eea
The general solution of \p{newsd} is an analytic function $\cE(a)$
depending on the single real $U(1)$ invariant variable $a=\nu\bnu$
which is quartic in the  auxiliary fields $V_\ab$ and $\bar{V}_{\da\db}$:
\bea
&& E^{ds}(\nu, \bnu) = \cE(a) = \cE(\nu\bnu)~,\;
\cE(0)=0~.
\label{condfin}
\eea

We come to the notable result that in the representation \p{legact} the 
{\it whole} class of nonlinear extensions of the Maxwell action
admitting the on-shell $SO(2)$ duality is parametrized by an arbitrary 
$SO(2)$ invariant real function of one argument $E^{ds} = \cE(\nu\bnu)$.
A remarkable property of $E^{ds}$ is that its power expansion collects 
only terms $\sim \nu^n\bnu^n$, i.e. those of $4n$-th order in the fields. 
Below we shall present this expansion for a few examples, including the 
notorious case of Born-Infeld theory.

It is evident that the bilinear part of the duality-symmetric Lagrangian 
in the $(F,V)$-representation \p{auxfree} is not invariant
\be
\delta_\omega \cL_2(F,V)=i\omega(F^2+2V^2-2F\cdot V-\mbox{c.c})=
i\omega(\vp-\bvp)~.
\ee
Thus the continuous $SO(2)$ duality in the $(F,V)$-representation amounts 
to a ``partial'' $SO(2)$ symmetry of the entire Lagrangian: it is a 
symmetry of its interaction part $E^{ds}(\nu, \bnu)$. It should be pointed 
out that the auxiliary field $V_{\alpha\beta},\bar V_{\dot\alpha\dot\beta}
$ is not subjected off shell to any constraint (as distinct from the 
Maxwell field strength which is subjected to the Bianchi identity), so the 
characterization of the $SO(2)$ duality-symmetric systems in the 
$(F,V)$-representation as those with the $SO(2)$ invariant self-interaction 
is valid {\it off shell}.

Let us consider the general $U(1)$ invariant interaction $\cE(\nu\bnu)$.
In order to construct the corresponding  Lagrangian $L(\vp,\bvp)$ one 
should solve the algebraic equations for $V_\ab(F)$ or $[V(F)]^2=
\nu(\vp,\bvp)$
\be
F_\ab=V_\ab(1+\bnu \cE_a)~\Rightarrow~\vp=\nu(1+\bnu \cE_a)^2~.\lb{FVsd}
\ee
Using eq. \p{FVsd}, one can derive the general equations relating the 
auxiliary variables $\nu$ and $a = \nu\bar\nu$  to the original variables 
$\vp, \bvp$
\bea
&\nu(1-a^2\cE^4_a)=\vp-\bvp a\cE_a^2+2a\cE_a(a\cE_a^2-1)\,,\nn& \\
& (1+a\cE_a^2)^2\vp\bvp=a[\cE_a(\vp+\bvp)+(1-a\cE^2_a)^2]^2~.\nn&
\eea
Note that it is not easy to find examples of the function $\cE(a)$ for 
which the algebraic equation for $\nu(\vp,\bvp)$ becomes explicitly 
solvable. In the next Section we shall consider an alternative choice of 
the auxiliary scalar variables which simplifies the explicit construction 
of duality-symmetric Lagrangians.

Finally, let us examine which restrictions on the interaction Lagrangian
$E(\nu, \bnu)$ are imposed by the requirement of the discrete
self-duality with respect to the exchange $F(A) \leftrightarrow F^\D(A^\D)
$. We shall do it in two ways.

We shall need a first-order representation of the Lagrangian \p{legact} 
analogous to \p{1stor}. Let us treat $\cL(V,F)$ in eq.\p{legact} as a 
function of two independent complex  variables $V_{\alpha\beta}, 
F_{\alpha\beta}$ and implement the Bianchi identities for $F_\ab, 
\bar{F}_{\da\db}$ (amounting to the expressions \p{defF}) in the 
Lagrangian via the dual field-strength $F^\D_\ab(A^\D)$ \p{PD}:
\bea
\widetilde{\cL}[V,F,F^\D] \equiv \cL(V,F) +\,i[F^\D\cdot F - 
\bar{F}^\D\cdot\bar{F}]~. \lb{modB}
\eea
The algebraic equation of motion for $V^\ab$, i.e. $\partial\widetilde{L}
/\partial V^\ab = 0$, is just the relation \p{FV}. On the other hand,
since $F_{\alpha\beta}, \bar F_{\dot\alpha\dot\beta}$ enter only bilinear 
part of the full Lagrangian in \p{modB}, varying \p{modB} with respect to 
$F_\ab$ (with keeping $V_{\alpha\beta}, \bar V_{\dot\alpha\dot\beta}$ 
off-shell) yields the exact linear relation
\be
F_\ab-2V_\ab=-iF^\D_\ab(A^\D) \quad (\mbox{and c.c.})   \lb{FVP}
\ee
as the corresponding equation of motion. As the result, one can 
explicitly find the dual form of \p{modB} in terms of $F^\D_\ab,
\bar{F}^\D_{\da\db}$ and $V_\ab, \bar V_{\da\db}\,$, expressing
$F_\ab$ and $\bar F_{\da\db}$ from eq. \p{FVP}:
\be
 \widetilde{\cL}[V,F(V,F^\D), F^\D]\equiv \widetilde{\cL}(U,F^\D)
 = \cL_2(U,F^\D)+ E(-u,-\bar{u})~, \label{mediate}
\ee
where
$$ U_\ab \equiv \Lambda V_\ab=-iV_\ab~,\q u=U^\ab U_\ab~.
$$

The discrete self-duality now amounts to demanding the Lagrangian 
\p{mediate} to have the same form in the variables $U, F^\D$ as the 
original Lagrangian $\cL (V, F)$ has in terms of $V, F$. Comparing the 
dual Lagrangian \p{mediate} with the original one \p{legact}, one firstly 
observes that $\cL_2$ in \p{mediate} looks the same in terms of the 
variables $U, F^\D$ as the original $\cL_2$, eq. \p{auxfree}, in terms of 
$V, F$. Then the necessary and sufficient condition of the discrete 
self-duality is the following simple restriction on the interaction 
function $E$ \cite{IZ}
\be
 E(\nu,\bnu)=E(-\nu,-\bnu)~. \lb{discr2}
\ee

Another proof is an analog of the on-shell consideration based on eqs.
\p{discrSO}, \p{Lambda}, \p{discinvar} in the standard formulation. Let us 
consider the transformation of $\cL(V,F)$ \p{legact} with respect to a 
discrete version of the $U(1)$ transformations \p{Vtrans}
\bea
&&\Lambda F_\ab=i(F_\ab-2V_\ab)=P_\ab~,\q\Lambda V_\ab=-iV_\ab~,
\lb{Vdiscr}\\
&&\Lambda \cL(V,F)=\cL_2(V,F)+E(-\nu,-\bnu)+iP\cdot F-i\bar P\cdot\bar F~.
\eea
By analogy with the condition \p{discinvar} the requirement of discrete
self-duality in the $(F,V)$-representation can now be reformulated as the 
$\Lambda$-invariance of the following function:
\be
I(F,V)=\cL(F,V)+{i\over2}P\cdot F-{i\over2}\bar P\cdot\bar F~.
\ee
We end up with the same condition \p{discr2} for $E(\nu, \bnu)$.

Obviously, an arbitrary $SO(2)$-invariant function $E^{ds}(\nu, \bnu) =
\cE(\nu\bnu)$ corresponding to a $SO(2)$ duality-symmetric system
automatically satisfies the discrete self-duality condition \p{discr2}.
This elementary consideration provides us with a simple proof of the fact
(mentioned in Sect. \ref{A}) that the $SO(2)$ duality-symmetric systems
constitute a subclass in the set of those revealing the discrete 
self-duality.

\setcounter{equation}0
\section{An alternative auxiliary field representation} \lb{D}

Eq. \p{FV} (or eqs. \p{algsc}) can be treated as an algebraic relation 
between two independent arguments of the function $\cL(F,V)$ \p{legact}.
Eliminating variables $F_\ab$ in this function, one can define an on-shell
$\nu$-representation of the general nonlinear Lagrangian
\be
L[\varphi(\nu, \bnu), \bar\varphi(\nu, \bnu)] \equiv
\hat{L}(\nu,\bnu) = E +{1\over2}\nu(E_\nu^2-
 2E_\nu-1)+{1\over2}\bnu(E_\bnu^2- 2E_\bnu-1)\,.
  \lb{hatLa}
\ee
However, this representation with $E=\cE(a)$ is not much helpful for
finding explicit examples of Lagrangians $L^{ds}(\vp,\bvp)$ in terms of 
the initial variables \p{compl}. It proves useful to define an alternative 
representation for the duality-symmetric Lagrangians,
$\tilde{L}(\mu,\bmu) \equiv \hat{L}[\nu(\mu, \bmu),\bnu (\mu, \bmu)])$, 
introducing new scalar auxiliary variables $\mu,\bmu$. Basic quantities of 
this $\mu$-representation are related to the corresponding quantities of 
the $\nu$-representation via the Legendre transformation. In Sect.\ref{C} 
we shall see that the defining algebraic equation of this
$\mu$-representation is more convenient for constructing explicit solutions 
of the $U(1)$ duality condition than the analogous one in the 
$\nu$-representation.

Let us introduce new complex scalar fields
\be
\mu(\nu,\bnu)=E_\nu~,\q \bmu(\nu,\bnu)=E_\bnu\lb{munu}
\ee
and consider the complex Legendre transformation $E(\nu,\bnu)\rightarrow
H(\mu,\bmu)$
\be
E(\nu,\bnu)-\nu E_\nu-\bnu E_\bnu=H(\mu,\bmu)~.
\lb{compLeg}
\ee
The corresponding inverse transformation is
\be
E(\nu,\bnu)=H(\mu,\bmu)-\mu H_\mu-\bmu H_\bmu \lb{compLeg1}
\ee
and
\be
\nu(\mu,\bmu)=-H_\mu~,\q \bnu(\mu,\bmu) = -H_\bmu~. \lb{nuK}
\ee

Note that the standard conditions \p{stcond} for the function $E(\nu,\bnu)$ 
do not imply any restriction on the second derivatives of this function.
However, for the transformed function $H(\mu,\bmu)$ to be analytic at the 
origin and, respectively, for the relation \p{munu}, \p{nuK} to be 
invertible, one is led to impose the following subsidiary condition on the 
Jacobian $J(\nu,\bnu)\equiv|E_{\nu\nu}|^2-|E_{\nu\bnu}|^2$ of the Legendre 
transformation:
\be
J(0,0) \neq 0~.\lb{Jacob}
\ee
It implies an analogous condition for $H(\mu, \mu)$ and selects those 
$L^{int}(\varphi, \bar\varphi)\,$, the $(\varphi, \bar\varphi)$-expansion 
of which starts with a non-degenerate 2nd order term. Below we shall limit 
our study to such analytic functions $H(\mu,\bmu)$.

Using eqs. \p{munu},\p{GE} and \p{exprG} one can find how $\mu$ is mapped 
on the derivative $L_\vp$
\be
\mu(L_\vp)=\frac{1+2L_\vp}{1-2L_\vp}=G^{-1}-1~,\q L_\vp=\frac{\mu-1}{2(\mu
+1)}~.\lb{muLvp}
\ee
The basic algebraic relation of the $\nu$-representation \p{algsc} can be 
transformed as follows
\be
\vp(\mu,\bmu)=-(1+\mu)^2H_\mu   \quad \mbox{(and c.c.)}~.\lb{vpmurel}
\ee
In order to find the corresponding Lagrangian $L(\vp,\bvp)$ one should
solve this basic relation for the function $\mu(\vp,\bvp)$. This solution 
can be analyzed perturbatively for any real analytic function $E$ (or $H$). 
However, the explicit solutions can be found only for some special cases.

Performing the Legendre transformation $E\leftrightarrow H$ in the 
Lagrangian \p{hatLa} (with the condition \p{Jacob} imposed) one can cast it 
in the $\mu$-representation
\be
\tilde{L}(\mu,\bmu)=\hat{L}[\nu(\mu,\bmu),\bnu(\mu,\bmu)]={1\over2}(1-
\mu^2)H_\mu+{1\over2}(1-\bmu^2)H_\bmu+H(\mu,\bmu)~.\lb{Dlagr}
\ee

It is interesting to note that this Lagrangian and the relation 
\p{vpmurel} can be reproduced from an off-shell Lagrangian with $\mu $ as 
an independent complex auxiliary field
\be
\tilde{\cL}(\vp,\mu)=\frac{\vp(\mu-1)}{2(1+\mu)}
+\frac{\bvp(\bmu-1)}{2(1+\bmu)}+H(\mu,\bmu)~.\lb{Lvm}
\ee
Indeed, varying \p{Lvm} with respect to $\mu$ one obtains just eq. 
\p{vpmurel}. Substituting the latter back in \p{Lvm}, one recovers the 
on-shell representation $\tilde{L}(\mu,\bmu)$, eq. \p{Dlagr}. The off-shell
Lagrangian  \p{Lvm} is an analogue  of the auxiliary-field reformulations
of the Born-Infeld Lagrangian  \cite{RT,Ts} (see Sect. \ref{C}).

Since the auxiliary fields $\nu$ and $\mu$ are related via the Legendre
transform \p{compLeg}, \p{compLeg1}, a similar off-shell Lagrangian should 
also exist for the on-shell $\nu$-representation \p{hatLa}, with the 
$\varphi \leftrightarrow \nu$ relation \p{algsc} arising as the appropriate 
algebraic equation of motion for $\nu$. However, the derivation of such a 
Lagrangian is not straightforward.

Let us turn to duality issues in the $\mu$ representation. Using eq.
\p{muLvp} and the formula  \p{delta1} for the variation $\delta L_\vp$,  
one can show that the $SO(2)$ duality group acts on $\mu $ as a linear 
$U(1)$ transformation
\be
\delta_\omega \mu=2i\omega\mu~.
\ee
Eq.\p{corol} implies the relation
\be
{1\over4}\vp(1-4L^2_\vp)=E_\nu\nu  =-\mu H_\mu~.
\ee
Then the $U(1)$ duality condition \p{sdI} in the $\mu$-representation is 
equivalent to the condition of $U(1)$-invariance of $H(\mu, \bmu)$
\bea
&&\delta_\omega H= 2i\omega(\mu H_\mu - \bmu H_\bmu) = 0~\Rightarrow~
H^{ds}(\mu,\bmu)=I(b)~,
\eea
where $I(b)$ is a real  function of the invariant argument $b=\mu \bmu$. 
The Jacobian condition \p{Jacob} now amounts to the one-dimensional 
relations
\be
\cE_a(0)\neq 0  \q \Leftrightarrow \q I_b(0) \neq 0\,.\lb{Icond}
\ee

Thus the  solution of the $U(1)$ duality condition has the following form 
in the $\mu$-representation:
\be
\nu(\mu,\bmu)=-\bmu I_b~,\q\vp= -(1+\mu)^2\bmu I_b\,.
\lb{phiR}
\ee
This solution is easily checked to provide the correct transformation rule 
for $\varphi $
\be
\delta_\omega\vp=2i\omega\vp\frac{\mu-1}{\mu+1}=4i\omega\,\vp L_\vp~.
\ee
From the definition of $\mu$ and the relations \p{phiR} it is 
straightforward to derive
\bea
&&a\equiv\nu\bnu=bI^2_b~,\lb{ERrel}\\
&&\mu=\bnu \cE_a~\Rightarrow~\frac{d\cE(a)}{da} =-\left(\frac{dI(b)}{db}
\right)^{-1}~.\lb{ERrel00}
\eea
The one-dimensional Legendre transform \p{compLeg}, \p{compLeg1} for the 
$U(1)$-invariant functions in the $\nu$- and $\mu$-representations reads
\be
\cE(a)=I(b)-2bI_b, \q I(b) = \cE(a) - 2a \cE_a\;. \lb{1dLeg}
\ee
Relations \p{ERrel}, \p{ERrel00} can be directly derived from \p{1dLeg}.

The general expression for the $\mu$-representation of the $U(1)$ 
duality-symmetric Lagrangian follows by substituting $I(b)$ for $H(\mu, 
\bmu)$ into $\tilde{L}(\mu,\bmu)$ defined by eq. \p{Dlagr}
\be
\tilde{L}^{ds}(\mu,\bmu)={1\over2}(\mu+\bmu)(1-b)I_b+I(b)
~.\lb{Klagr}
\ee
It possesses the correct $U(1)$ transformation properties
\bea
&&\delta_\omega \tilde{L}^{ds}=i\omega(\mu-\bmu)(1-b)I_b=i\omega(\vp-
\bvp)\nn\\
&&\delta_\omega(\vp-\bvp)=4i\omega(\tilde{L}^{ds}-I)~.
\eea
Eq. \p{Lvm} with the substitution $H\rightarrow I(b)$ provides an off-shell
description of the considered restricted class of the $U(1)$
duality-symmetric theories in the $\mu$ representation. Using eqs. 
\p{muLvp} and \p{phiR} one can find
\be
\vp L_{\vp} = {1\over 2}(1-\mu^2)\bmu\,I_b
\ee
and, substituting this in \p{Klagr}, show that the analytic functions 
$I(b)$ satisfying the condition \p{Icond} constitute a particular class of 
the invariant functions $I(\vp, \bvp)$ defined in \p{invar}:
$$
I[b(\vp,\bvp)]=L(\vp,\bvp)-\vp L_\vp-\bvp L_\bvp\equiv I(\vp,\bvp)~.
$$
This class of functions $I(\vp, \bvp)$ is characterized by the presence of
nonzero term of 4th order in Maxwell field strength in their 
$(\vp,\bvp)$-expansion,  $ I(\vp, \bvp) = I_b(0)\,\vp\bvp + \ldots $.

Note that the general rescaling \p{rescal} which preserves the $SO(2)$ 
duality corresponds in the $\mu$-representation to the rescaling
$I(b)\rightarrow rI(b)$. In particular, \p{mirr} corresponds to the
reflection $I(b) \rightarrow -I(b)$. In Sect.\ref{C} we shall consider the 
impact of this reflection on the Lagrangian of the Born-Infeld theory.

The basic algebraic problem of the $\mu$-representation is to restore the
function $\mu(\vp,\bvp)$ and then $b(\vp, \bvp)$ by the given inverse 
function $\vp(\mu,\bmu)$ in \p{phiR}. Once the latter function is analytic, 
the analyticity of $\mu(\vp,\bvp)$ is guaranteed by the implicit function 
theorem. The basic algebraic equation for the function $b(\vp,\bvp)\,$,
\be
(b+1)^2\vp\bvp= b[(\vp+\bvp)-I_b(b-1)^2]^2\lb{Jphi}\,,
\ee
and  the corresponding representation for $\mu(\vp,\bvp)\,$,
\be
\mu=-\frac{\bvp-b\vp-2b(b-1)I_b }{I_b(1-b^2)}\,, \lb{muphi0}
\ee
follow from eqs. \p{phiR}. In the next Section we shall see that the 
relations \p{Jphi} and \p{muphi0} are helpful while seeking the explicit 
solutions of the $SO(2)$ duality constraint.

Finally, let us analyze  the discrete duality of $\tilde{L}(\mu,\bmu)$
\p{Dlagr}. In the $\mu$-representation, the appropriate discrete 
transformations are
\be
\Lambda \mu(\nu,\bnu)\equiv \mu(-\nu,-\bnu)=-\mu(\nu,\bnu)~.
\ee
The discrete self-duality of $\tilde{L}(\mu,\bmu)$ is then equivalent
to the symmetry
\be
\Lambda H(\mu,\bmu)\equiv H(-\mu,-\bmu)=H(\mu,\bmu)~,
\ee
which guarantees the correct $\Lambda$ transformation \p{Lambda} of the 
full Lagrangian \p{Dlagr}.

\setcounter{equation}0
\section{Examples of duality-symmetric systems}\lb{C}

\subsection{Born-Infeld theory}
The Lagrangian of the Born-Infeld theory has the following form in terms 
of complex invariants \p{compl}
\bea
&& L(\vp,\bvp) =
\left[1-Q(\vp,\bvp)\right]~, \lb{biact}
\eea
where
\bea
Q(\vp,\bvp) = \sqrt{1+X}~,\q
X(\vp , \bvp) \equiv
(\vp+\bvp)+(1/4)(\vp-\bvp)^2~.\lb{not}
\eea
The power expansion of the BI Lagrangian is
\bea
L=-{1\over2}(\vp+\bvp)+{1\over2}\vp\bvp
-{1\over4}\vp\bvp(\vp+\bvp)
+{1\over8}\vp\bvp(3\vp\bvp+
\vp^2+\bvp^2)+O(\vp^5)~. \lb{BIpert}
\eea

In the BI theory the function  \p{defP} has the following explicit form
\bea
P_\ab(F)=i\frac{\partial L}{\partial F^\ab}
=-iF_\ab Q^{-1}(\vp,\bvp)[1+{1\over2}(\vp-\bvp)]
\eea
and the basic $U(1)$-transformations of the scalar variable is
\be
\delta_\omega\vp=-2i\omega\frac{\vp[1+{1\over2}(\vp-\bvp)]}{Q}~.
\ee
The function $G(\vp, \bvp)$ relating the variables $V_\ab$ and $F_\ab$
and defined by eq. \p{exprG}, is given by the expression
\bea
&&G={1\over 2}\left\{1 + Q^{-1}
\left [1 + {1\over2}(\vp - \bvp)\right]\right\}~. \lb{Gbi}
\eea

Let us first discuss the $\mu$-representation of BI theory. It is easy to 
find the  relations
\bea
&& \vp=\frac{2\bmu(1+\mu)^2}
{(1- \mu\bmu)^2}~, \q  \bvp=\frac{2\mu(1+\bmu)^2}
{(1- \mu\bmu)^2}~,\lb{zG}
\eea
which correspond to the following choice of the invariant function in
the $\mu$-representation \p{phiR}:
\be
 I(b)=\frac{2b}{b-1}\,, \q I_b = -\frac{2}{(b-1)^2} \lb{RBI}
\ee
(with $b = \mu\bmu$). Using this choice of the auxiliary function in 
\p{Jphi} we obtain the quadratic equation for the invariant variable $b$
\be
\vp\bvp\, b^2+[2\vp\bvp-(\vp+\bvp+2)^2]\,b+\vp\bvp=0~.\lb{BIb}
\ee
The invariant and linearly transforming  functions $b$ and $\mu$ obtained 
by solving \p{BIb} and using the general formulas \p{muLvp} or \p{muphi0} 
are given by the expressions
\bea
b=\frac{\vp\bvp}{[1+Q+{1\over2}(\vp+\bvp)]^2}~, \q
\mu= G^{-1}-1=\frac{Q-1-{1\over2}(\vp-\bvp)}{Q+1+{1\over2}(\vp-\bvp)}~.
\lb{KBI}
\eea
The corresponding representations for the off- and on-shell BI Lagrangian 
read
\bea
\tilde{\cL}(\vp,\mu)=\frac{2b}{b-1}+\frac{\vp(\mu-1)}{2(1+\mu)}
+\frac{\bvp(\bmu-1)}{2(1+\bmu)}~, \q \tilde{L}(\mu,\bmu)=\frac{\mu+\bmu
+2b}{b-1}~.\lb{Kblag}
\eea

Note that  the authors of Ref. \cite{RT} considered a polynomial
off-shell representation of the BI Lagrangian with two complex auxiliary 
fields. The  basic auxiliary field $\chi$ of this representation is 
related to our fields $\vp,\mu$ and $b=\mu\bmu$ as follows
\bea
\chi+{1\over2}\chi\bar\chi=\vp~,\q \chi=\frac{2(\mu+b)}{b-1}~,\q
\tilde{L}(\mu,\bmu)=-{1\over2}(\chi+\bar\chi)~.
\eea

Let us also study the original $(F,V)$-representation for the BI case.
Our aim is to find $\cE(a)$ as a function of the variable
\bea
&&  a=\nu\bnu=\frac{4b}{(b-1)^4}
\lb{xG}
\eea
(recall eq. \p{ERrel}). Introducing $t \equiv (b-1)^{-1}$, one finds that 
$t$ satisfies the following quartic equation:
\be
 t^4 + t^3-{1\over4} \nu\bnu = 0~, \; t(\nu = \vp =0) = -1~. \lb{quart}
\ee
It allows one to express $t$ in terms of $a \equiv \nu\bnu\,$:
\be
 t(a)=-1-\frac{a}{4}+\frac{3a^2}{16}-\frac{15a^3}{64}+
\ldots~.
\ee
One can easily write a closed expression for $t(a)$ as the proper solution
of \p{quart}, but we do not present it here.

Now we are ready to find the invariant self-interaction $\cE(\nu\bnu)$ for
this case. Taking into account Eqs.\p{RBI} and \p{1dLeg} we find a simple 
expression for the self-interaction through the real variables $b$ or 
$t(a)$ (see also\cite{IZ})
\bea
\cE_{\B\I}[a(b)]=\frac{2b(1+b)}{(1-b)^2}=2[2t^2(a)+3t(a)+1] =
\frac{a}{2}-\frac{a^2}{8}+ \frac{3a^3}{32}+\ldots~.
\lb{explE}
\eea

It is easy to show that \p{Jphi} is reduced to quadratic equation only
for the one-parameter family of functions
\be
I_b=-2r/(b-1)^2\,, \lb{RBI1}
\ee
which corresponds to performing the transformation \p{rescal} in the BI 
Lagrangian \p{biact}. A new duality-symmetric Lagrangian (not reducible to 
the BI one) is obtained in the case $r=-1$
\bea
L^{(-)}=-1+\sqrt{1-\vp-\bvp+{1\over4}(\vp-\bvp)^2} =-1+\sqrt{1+{\bf E}^2-
{\bf B}^2-({\bf E}{\bf B})^2}~,
\eea
where ${\bf E}$ and ${\bf B}$ are electric and magnetic fields, 
respectively. This Lagrangian is obtained from the BI one \p{biact} by 
changing its overall sign and making the replacement $F_{mn} \rightarrow 
\tilde{F}_{mn} = {1\over 2}\epsilon^{mnkl}F_{kl}$,
or
\be
{\bf E}\rightarrow {\bf B}~,\q{\bf B}\rightarrow -{\bf E}\,.
\ee
It would be interesting to find out the physical meaning and implications
of this ``magnetic'' counterpart of the BI theory.

\subsection{Exact duality-symmetric Lagrangians corresponding to \break 
solvable algebraic equations}
The initial data for restoring  Lagrangians $L^{ds}(\varphi, \bar\varphi)$
by the known function $I(b)$ in the $\mu$-representation is the equation 
\p{Jphi} for $b$ and the following representation for the Lagrangian 
\p{Klagr}
\be
\tilde{L}^{ds} = -{1\over 2}\left(\frac{1-b}{1+b}\right)\left[\varphi + 
\bar\varphi + 4bI_b(b)\right]+ I(b)\,, \lb{Lagrb}
\ee
which is obtained by substituting the expression \p{muphi0} for $\mu$ into
\p{Klagr}. The key idea of finding out new explicit examples of
duality-symmetric models is to pick up those $I(b)$ for which the basic 
eq. \p{Jphi} is simplified as much as possible. Below we analyze several 
examples of the  function $I(b)$ which make \p{Jphi} a solvable algebraic
equation for $b(\vp,\bvp)$.

As already mentioned, the quadratic equation is obtained only in the case 
of BI theory \p{RBI} and its ``magnetic'' counterpart \p{RBI1} (with 
$r=-1$).

Next in complexity is the following  ansatz for $I_b$
\be
I_b=-\frac{2-cb}{(1-b)^2}  \q \Rightarrow \q I(b) = (c-2)\frac{b}{1 -b} + 
c\,\ln (1-b)\,, \lb{3rdor}
\ee
where $c$ is some constant and we employed the conditions $I(0) = 0,~
I_b(0)=-2$. Being substituted into \p{Jphi}, this ansatz  gives the cubic 
algebraic equation for the unknown $b(\vp,\bvp)$
\be
c^2b^3-b^2[4c+2c(\vp+\bvp)+\vp\bvp]+b[(\vp+\bvp+2)^2-2\vp\bvp]-\vp\bvp
=0~.\lb{Jc}
\ee
It is straightforward to write down the explicit solution of this equation 
for $b(\vp,\bvp)$ as the appropriate analytic function of $\varphi, 
\bar\varphi$ (vanishing at $\vp = \bvp =0$) and to find the precise 
expression for the related $SO(2)$ duality-symmetric Lagrangian by 
substituting this solution into \p{Lagrb}.

The case $c=0$ yields the BI theory, while for any other value of $c$
we obtain new examples of the duality-invariant systems. In the special 
case $c=2$, $I_b$ and $I$ in eq.\p{3rdor} are simplified to $I_b=
-2(1-b)^{-1}, \;I(b) = 2\ln (1-b)$. With this choice, the relation between 
different auxiliary-fields representations is also simplified
\bea
&&a=\frac{4b}{(b-1)^2}~,\q b(a)=\frac{a+2-2\sqrt{1+a}}{a}~, \nn \\
&&\cE(a)=2(\sqrt{1+a} -1)-2\ln{1\over2}(1+\sqrt{1+a})\,.
\eea

A different ansatz for $I_b$ also leading to a comparatively simple 
algebraic equation for $b$ is as follows
\be
I_b=-\frac{2\sqrt{1-cb}}{(b-1)^2}~.
\ee
Eq. \p{Jphi} is reduced to
\be
(b+1)^2\vp\bvp-b(\vp+\bvp)^2-4b(1-cb)=4(\vp+\bvp)b\sqrt{1-cb}\,,
\lb{Jphi00}
\ee
which is equivalent to a quartic equation. In the limit $c\rightarrow 0$ 
eq. \p{Jphi00} becomes quadratic and one recovers the BI theory.

One more solvable ansatz for $I_b$ is
\be
I_b=-2\frac{1}{(1-cb)(b-1)^{2}}\,.
\ee
It reduces \p{Jphi} to the following quartic equation:
$$
(1-cb)^2(b+1)^2\vp\bvp=b[(1-cb)(\vp+\bvp)+2]^2~.
$$

\subsection{Discrete duality examples}

Let us firstly consider, directly in the original $\varphi, \bar\varphi$
representation, two simple examples of the Lagrangians exhibiting
discrete self-duality.

The first example is the Lagrangian which depends on a single real variable 
$\phi=\vp+\bvp$
\be
L=1-\sqrt{1+\phi}~.
\ee
Another example is the holomorphic nonlinear Lagrangians
\be
L_h(\vp,\bvp)=l(\vp)+\bar l(\bvp)~, \q l(\vp)=1-\sqrt{1+\vp}~.\lb{holom1}
\ee
It is a straightforward exercise to check that both these Lagrangians
respect self-duality under Legendre transformation as it was defined in
Subsect. 2.2. At the same time, they are not $SO(2)$ duality-symmetric.

Two other examples of systems with a discrete duality can be introduced 
in the framework of $\nu$-representation. The first one corresponds to 
the choice
\be
E={1\over2}N^2~,\q N=\nu+\bnu~,
\ee
then after eliminating auxiliary fields the final Lagrangian $L(\varphi, 
\bar\varphi)$ is a function of the single real variable $\phi=\vp+\bvp$. 
The basic algebraic equation is cubic
\be
\phi=N(1+N)^2\,,\lb{phN}
\ee
and it can be solved in radicals
\bea
&&N(\phi)=-{2\over3}+A_+(\phi)+A_-(\phi)=\phi-2\phi^2+7\phi^3-24\phi^4
+\ldots~, \\
&&A_\pm(\phi)=\sqrt[3]{{1\over27}+{\phi\over2}\pm\sqrt{B(\phi)}}\,,\qquad
B(\phi)={1\over27}\phi+{1\over4}\phi^2\,.\nn
\eea
Despite the presence of radical $\sqrt{B(\phi)}$ in $A_\pm(\phi)$,
the function  $N(\phi)$ is analytic.

More complicated Lagrangian corresponds to the following choice
of the even self-interaction function $E(\nu, \bnu) = E(-\nu, -\bnu)$:
\bea
&&E(\nu,\bnu)={1\over2}(\nu^2+\bnu^2)~,\q E_\nu=\nu=\mu~,\\
&&L^{int}(\vp,\bvp)={1\over2}(\nu^2+\bnu^2)+\nu^3+\bnu^3~.\nn
\eea
The holomorphic algebraic equation
\be
 \vp=\nu+2\nu^2+\nu^3
\ee
can be explicitly solved similarly to eq. \p{phN}. The corresponding 
Lagrangian $L(\varphi, \bar\varphi)$ is holomorphic like \p{holom1}.

\section{Conclusion}
We introduced a new $(F,V)$-representation for the Lagrangians of 
nonlinear electrodynamics and showed that it provides a simple 
description of systems exhibiting the properties of $U(1)$ duality or/and  
discrete self-duality in terms of real function of auxiliary bispinor 
complex fields, $E(V, \bar V)$. This function encodes the entire 
self-interaction in the $(F,V)$-representation. The duality properties 
prove to be related to some linear off-shell symmetries of this general 
function $E$. We also defined an alternative $\mu$-representation
and demonstrated its convenience and efficiency for constructing new 
explicit examples of duality-symmetric Lagrangians.

The auxiliary linearly transforming variables have also been used to
construct the general solution of the $U(n)$ duality constraint for the 
interaction of $n$ Abelian gauge fields in \cite{IZ1}. The generalization 
to the $U(n)$ case is a straightforward extension of the formalism 
described above, so we do not present it here and send the interested 
reader to Ref. \cite{IZ1}.

It is the interesting task to extend our consideration to the case of 
$N=1$ and $N=2$ supersymmetric extensions of nonlinear electrodynamics 
\cite{KT} in order to obtain a general characterization of the 
corresponding duality-symmetric systems. One more urgent problem is to 
define an analog (if existing) of the $(F,V)$- representation for 
non-Abelian BI theory and its superextensions. It could shed more light 
on the structure of these theories which have deep implications
in string theory and still remain to be completely understood.

\section*{ Acknowledgements}

 This work was partially supported by INTAS grant No 00-254,
RFBR grant No 03-02-17440, RFBR-DFG grant No 02-02-04002, grant DFG
436 RUS 113/669, and a grant of Heisenberg-Landau Programme.

\end{document}